 \documentstyle[preprint,aps,eqsecnum]{revtex}

\def\beq#1{\begin{equation} \label{#1}}
\def\eeq{\end{equation}}
\newcommand{\bea}{\begin{eqnarray}}
\newcommand{\eea}{\end{eqnarray}}
\def\bra#1{\left\langle #1\right\vert}
\def\ket#1{\left\vert #1\right\rangle}
\def\epsp{\epsilon^{\prime}}
\def\NPB{{ Nucl. Phys.} B}
\def\PLB{{ Phys. Lett.} B}
\def\PRL{ Phys. Rev. Lett.}
\def\PRD{{ Phys. Rev.} D}
\def\AJP{{\em Am. J. Phys.}}
\begin{document}
\tighten


\title{Directed Spontaneous Emission from $N$-atom Extended Ensemble}

\author{HARRY J. LIPKIN}

\address{Department of Particle Physics
  Weizmann Institute of Science, \\ Rehovot 76100, Israel}

\address{School of Physics and Astronomy,
Raymond and Beverly Sackler Faculty of Exact Sciences,\\
Tel Aviv University, Tel Aviv, Israel}

\address{Advanced Photon Source, Argonne National Laboratory,\\
Argonne, IL 60439-4815, USA}


\def\beq#1{\begin{equation} \label{#1}}
\def\eeq{\end{equation}}
\def\bra#1{\left\langle #1\right\vert}
\def\ket#1{\left\vert #1\right\rangle}
\def\epsp{\epsilon^{\prime}}
\def\NPB{{ Nucl. Phys.} B}
\def\PLB{{ Phys. Lett.} B}
\def\PRL{ Phys. Rev. Lett.}
\def\PRD{{ Phys. Rev.} D}
\def\AJP{{\em Am. J. Phys.}}


\maketitle
\vskip + 2cm

\begin{abstract}%
Coherence and interference play crucial roles in emission and absorption of photons to and from large systems with many atoms.  Confusion has arisen because nuclear X-ray physicists and atomic quantum-optics physicists do not understand one another's individual
descriptions of related phenomena. Basic physics same for all wave lengths from optical to nuclear gamma ray photons.  But different languages are used to describe this physics in different domains. Crucial parameters vary over many orders of magnitude and what is intuitive or counterintuitive varies widely. Differences in parameters arising from differences between coherent emission effects in different domains produce very different results. Unified general treatment of the entire photon spectrum makes basic physics intelligible to all. In the optical region the mean distance between the scattering atoms is much longer than the photon wave length, Dicke superradiant scattering is isotropic and multiple scattering, Fano couplings are important and the lifetimes of intermediate stats are sufficiently short to be negligible. In X-ray scattering the mean distance between atoms is comparable to the photon wave length, Dicke superradiance is concentrated in a forward peak, multiple scattering and Fano effects are negligible, lifetimes are measurably long, speedup shortening the lifetime is important and most of the radiation is not elastically scattered but lost to absorption. Explicit calculations for a one-dimensional array shows the great difference between the case where the photon wave length is much shorter or comparable to the distance between nearest neighbors. A full investigation of the angular distribution and speedup of the intensity for two and three-dimensional scatterers give very different results for the two cases.
\end{abstract}

\vfill\eject

\renewcommand{\thefootnote}{\fnsymbol{footnote}}
\setcounter{footnote}{1}

\section {Introduction}

The paper \cite{Scully} ``Directed Spontaneous Emission from an Extended
Ensemble of $N$ atoms: Timing is Everything" considers a state prepared by
absorbing a photon of wave vector $\vec k_0$ in which one atom is excited and
we don't know which one. It asks ``When the absorbed photon is spontaneously
emitted will it go in 4$\pi$ sr. or will it be directionally correlated with
$\vec k_0$? The perhaps counterintuitive answer is the latter."

Such states have been prepared with photons of any wave length from
optical photons to nuclear gamma ray photons. The basic physics is the
same for all, but different languages are used to describe this physics
in different domains. Crucial parameters vary over many orders of
magnitude and what is intuitive and what is counterintuitive vary widely.
Confusion has arisen because nuclear X-ray physicists and atomic
quantum-optics physicists do not understand one another's individual
descriptions of related phenomena. As Maurice Goldhaber\cite{Maurice}
remarked about the discovery of the M\"ossbauer effect, ``This will teach
many old things to new people."

The purpose of this paper is to provide a unified description
which makes the basic physics intelligible to all, while pointing out
some important differences between coherent emission effects in different
domains and showing how differences in parameters produce very different
results.

The first time this question was considered and called ``superradiance" by
Dicke\cite{Super} the emitted photon did go in 4$\pi$ sr.  In the nuclear gamma
ray case this question has been investigated in detail both
theoretically\cite{HanTram,HJLIP} and experimentally\cite{Ruffer,vanBurck}. And
the latter answer above is not counterintuitive at all. And in other domains
Dicke  superradiance\cite{ctlee} gave different answers.

The general situation can be summarized as follows:

\begin{enumerate}

\item The angular distribution of the emitted photon depends upon upon the
parameters of the source. These parameters can vary by many orders of magnitude
between optical and nuclear $\gamma$ ray sources. The angular distribution of
the emitted photon can vary from isotropic to  strongly forward peaked.

\item The photon emission is always speeded up for a three-dimensional system.
The lifetime of this coherent or superradiant state is always much shorter than
the lifetime of an isolated atom or nucleus in the given excited state. But the
degree of speedup; i.e. the ratio of the lifetime of the superradiant state to
the single atom lifetime depends upon the parameters of the source.

\item The probability that the initial excitation will produce a photon with the
same wave length and energy as the initially absorbed photon varies widely over the
different domains. In quantum optics, this probability is of order unity; in
nuclear gamma radiation it is at best only a few percent and maybe considerably
smaller.

\item The probability that a nuclear excitation will produce a photon with the
same wave length and energy as the initially absorbed photon is very small for
the following reasons:

\begin{itemize}

\item The dominant process for energy transfer from an excited nucleus is
internal conversion, in which no photon is emitted and an electron is ejected
from one of the atomic shells. The ratio of the probability of this electron
ejection to the probability that a photon is emitted is called the internal
conversion coefficient and is greater than ten for most transitions relevant
here.

\item In photon emission from a nucleus that is free and at rest, energy is lost
in the recoil of the nucleus which balances the momentum of the emitted photon
and there is no photon emitted with the energy of the initially absorbed photon.

\item In photon emission from a nucleus that is in thermal equilibrium at a
sufficiently high temperature, the energy spectrum includes nuclei with
sufficient energy to compensate for the recoil energy. Only these can be
absorbed resonantly by another nucleus.

\item In photon emission from a nucleus bound in a crystal the recoil energy can
be exchanged with emission and absorption of lattice phonons. The probability of
elastic emission with no energy transfer to the lattice is called the
Debye-Waller or Lamb-M\"ossbauer factor. Only these elastically emitted photons
can participate in Dicke superradiance.

\end{itemize}

\item If the source dimension is not sufficiently small to allow the emitted
photon to escape without further interactions complicated rescattering effects
can occur which depend strongly on the source parameters and can vary wildly
between different systems and different wave lengths. These are very different
for optical photons and nuclear $\gamma$ rays as listed below. There is no
further discussion of multiple scattering in this paper.

\begin{itemize}

\item Continous rescattering of optical photons is called ``Fano coupling" and
plays an important role on optical superradiance

\item Rescattering of superradiant nuclear $\gamma$ rays can occur as the
$\gamma$ ray proceeds forward in a thick crystal. However because the speeded up
superradiant photons have a much wider energy spectrum than the natural line
width the absorption through the crystal distorts the energy spectrum, first
making a hole as photons within the natural line width are absorbled.
These effects are not
discussed in this paper.
\end{itemize}

\end{enumerate}

\section{Comparison of forward and $90^o$ emission from an ordered array}

The essential differences are seen in comparing forward and $90^o$ emission
from  an ordered array of $N$  atoms equally spaced by a distance $\vec D$.
Consider a state  prepared  by absorbing a photon of wave vector $\vec k_0$ in
which one atom is excited and we don't know which one.

In forward emission the path difference $\vec D$ between initial excitations of
neighboring  atoms in  the array is exactly compensated by the path difference
of the emitted photon.  Thus all the forward amplitudes are in phase and  the
total  forward amplitude is proportional to $N$.
\beq{isotrop}
A(0^o) = \sqrt {N}a; ~ ~ ~ I(0^o) = Na^2
\end{equation}
In $90 ^o$ emission there is  no path difference between emitted photons.
The phase difference $\vec k_0 \cdot \vec D$ between the initial excitations
of nearest neighbors thus remains in the outging radiation. The amplitudes with
this phase difference are easily summed in a geometric series to give
\beq{geomet}
A(90^o) = {{a}\over{\sqrt {N}}} \cdot \sum_{n=0}^{N-1} e^{in\vec k_0 \cdot \vec D}
  = {{a}\over{\sqrt {N}}}\cdot
{{e^{iN\vec k_0 \cdot \vec D}-1}\over{e^{i\vec k_0 \cdot \vec D}-1}}
\end{equation}
where $a$ denotes the amplitude emittted from a single atom and is assumed to
have an isotropic angular distribution
\beq{geomet2}
{{A(90^o)}\over{A(0^o)}} ={{1}\over{N}}\cdot
{{e^{iN\vec k_0 \cdot \vec D}-1}\over{e^{i\vec k_0 \cdot \vec D}-1}}
\end{equation}  
When  $\vec k_0 \cdot \vec D \approx 0$, $e^{iN\vec k_0 \cdot \vec D} \approx
1$, $A(90^o)\approx A(0^o)$
and the angular distribution is approximately isotropic.

But when $\vec k_0 \cdot \vec D$ is of order unity, the amplitude at
$90^o$ is strongly suppressed.
\beq{geomet3}
{{A(90^o)}\over{ A(0^o)}} \approx {{1}\over{N}}; ~ ~ ~
{{I(90^o)}\over{ I(0^o)}} \approx {{1}\over{N^2}}$$
\end{equation}
In a disordered linear array the intensity is the sum of intensities from N
nuclei with random phases.
\beq{geomet4}I_{dis}(90^o) = a^2; ~ ~ ~
{{I_{dis}(90^o)}\over{I(0^o)}} \approx {{1}\over{N}}
\end{equation}
Three different domains of values of the parameter
$\vec k_0 \cdot \vec D$ lead to very different conclusions.
\begin{enumerate}
\item
When  $N \vec k_0 \cdot \vec D \approx 0$, $e^{iN\vec k_0 \cdot \vec D} \approx
1$, $A(90^o)\approx A(0^o)$
and the angular distribution is approximately isotropic. This is simple Dicke
superradiance.
\item
When $\vec k_0 \cdot \vec D$ is of order unity,
$A(90^o) \approx (1/N)\cdot A(0^o)$.
The emission at $90^o$ is strongly suppressed. This is the case in nuclear
resonance scattering.
\item
When $0 \leq n\vec k_0 \cdot \vec D \leq 1$ for small but
finite values of $n$ much less than $N$ and
$ n \vec k_0 \cdot \vec D$ reaches unity for a finite value of $n \ll N$. Then
$ 0 \ll A(90^o) \ll A(0^o)$.
There may be interesting physics in this intermediate domain which may be
considered counterintuitive in quantum optics. It goes beyond simple isotropic
Dicke superradiance and does not exist in the X-ray region.

\end{enumerate}

\section {Explicit calculation of the intensity summed over the angular
distribution}

Forward coherence occurs only in the exact ${\vec k_0}$
direction. That the amplitude for the emitted photon to be
enhanced by a factor N in the exact forward direction can be shown to be true
in general for all sources. However, the exact forward direction has zero
solid angle. The forward intensity observed in an experiment is an integral
over a finite solid angle.

If individual sources of a system of N atoms are at positions denoted by
$r_\mu$ and
radiate isotropically the total
radiation intensity $I$ integrated over all angles can be shown to be

\beq{itot}
I_{tot}\equiv \int I d\Omega =
        {{4\pi}\over {N}} \cdot  \sum_\mu\,\sum_\nu i_1 e^{i(\phi_\mu -
\phi_\nu)} \cdot {{\sin(k|r_\mu - r_\nu|)}\over {k|r_\mu - r_\nu|}}
\end{equation}
where $i_1$
is intensity per unit solid angle for the transition from a single
excited nucleus and the phases $\phi_\mu$ and
$\phi_\nu$ depend upon the preparation of the source.
For simplicity we drop the subscript $0$ and define $\vec k \equiv \vec k_0 $
If the source is concentrated in a very small volume,
$k|r_\mu - r_\nu| \ll 1$, $I_{tot}\propto N$
and typical Dicke superradiance is observed.

If, however, $k|r_\mu - r_\nu| \gg 1$,
the factor $k|r_\mu - r_\nu|$
in denominator suppresses contributions from pairs
separated by large distances
and $I$ is no longer $\propto N$.

Explicit $N$ dependence is seen by assuming optimum phase, introducing
cylindrical co-ordinates $(r,z)$ for an axially symmetric  cylindrical
system with radius $a$ and length $L$,
evaluating the double sum by converting one sum to an integral, and
neglecting differences between the distances to boundaries of different
atoms.
\beq{cylind}
I_{tot}= 4\pi i_1 \int_0^L dz
\int_0^a 2 \pi r \rho dr
\cos (kz) \cdot {{\sin kR }\over {kR}}
\end{equation}
where $\rho$ denotes density of atoms and we define
\beq{defr}
R \equiv \sqrt{r^2+ z^2}; ~ ~ ~ R_{max} \equiv \sqrt{a^2+ z^2}
\end{equation}
Changing variables for the integration over $r$ and noting that
$rdr = RdR$ gives
\beq{itot2}
I_{tot}
= 8 \pi^2 i_1  \rho \int_0^L dz \int_z^{R_{max}}dR
\cos (kz) \cdot {{\sin kR }\over {k}}
\end{equation}

\beq{itot3}
I_{tot}= 8 \pi^2 i_1  \rho \int_0^L dz
\cos (kz) \cdot {{\cos kz - \cos kR_{max}}\over {k^2}}
\approx 8 \pi^2 i_1  \rho \int_0^L dz {{\cos^2 kz}\over {k^2}}
\approx 4 \pi^2 {{i_1  \rho L}\over {k^2}}
\end{equation}
where we have neglected the oscillating part of the integral over $z$.

This result can be compared with
the exact forward intensity which is always enhanced by a factor
$N=\pi a^2 L\rho$. This would give a total intensity of  $4\pi^2 i_1 a^2 L\rho$
if this intensity remained constant over the full $4\pi$ solid angle.
We can therefore define an ``effective solid angle" for the forward beam as
\beq{Omegaeff}
\Omega_{eff}= \frac{I_{tot}}{4\pi^2 i_1 a^2 L\rho}=\frac{1}{k^2 a^2}
\end{equation}
The result shows enhancement or ``speedup" of
intensity compared to  decay of a
single nucleus by a factor $2\pi \rho {{L}\over {k^2}} $
and a narrowing of the solid angle of the forward peak by a factor of
$k^2 a^2$. This shows that increasing the length of the source in the direction
of the incident photon increases the speedup as well as increasing the forward
intensity, while increasing the size of the source in the direction normal to
the incident photon direction does not increase the speedup; it only increases
the forward intensity at the expense of narrowing the angular distribution.

One particularly interesting case is the ``end fire mode" for a long thin
needle\cite{ctlee} of length $L$ and diameter $a$. The enhancement is seen to be given
is given by the length of the needle $L$.

Another interesting case is a flat disc with a thickness of only a few atomic
layers. The speedup is only  due to the number of atomic layers and independent
of the size of the disc. There is no enhancement for a two-dimensional disc.
Thus the enhancement is seen to arise only for a three-dimensional source.
No additional enhancement is found for one or two dimensional sources.

This can also be seen by explicit examination of the summation (\ref{itot})
for the one and two dimensional cases. In one dimension the sum vanishes for
large  values of $|r_\mu - r_\nu|$ and the periodic variation of the integrand
further  prevents any enhancement at large distances. The summation in two
dimensions does not vanish for large values of $|r_\mu - r_\nu|$ because the
$|r_\mu - r_\nu|$ denominator is canceled by a similar factor in the
two-dimensional integrand. However the periodic variation of the integrand
remains and still prevents any enhancement at large distances.

The enhancement can be seen as resulting from competing factors of enhancement
by a factor $N$ in the exact forward direction and reduction in the angular
width of the forward peak.

The forward enhancement can  be expressed as a
factor proportional to $R^d$ where $R$ is the linear size of the source and $d$
is the dimension. If the width scales like 1/R, we find that there is an
overall enhancement by a factor $R$ or $N^{(1/3)}$ for a three dimensional
spherical source and no similar enhancement for linear or planar  sources.

\section{Some further questions}

The question has been asked whether the quantum-optical analysis in the present
paper\cite{Scully} adds any new knowledge in a different language to what has
been long known in X-ray physics. In this example we note that X-ray physics is
defined by the region in which $\vec k_0 \cdot \vec D \geq 1$. The region where
$\vec k_0 \cdot \vec D \approx 0$ gives nearly isotropic Dicke superradiance and
is outside the domain of X-ray physics.

Another region explored in this paper\cite{Scully} does not exist in X-ray
physics; namely where $\vec k_0 \cdot \vec D$ is small but finite and $ n \vec
k_0 \cdot \vec D$ reaches unity for a finite value of $n \ll N$. There may be
interesting physics in this domain which goes beyond simple isotropic Dicke
superradiance and does not exist in the X-ray region. If so, it  should be
stated in a way that pinpoints the new  knowledge and can be understood by all
physicists and not only by a group  which speaks in its own jargon.

Some of the different aspects of coherence from ensembles with a single
excitation can be summarized as follows:

\begin{enumerate}

\item In forward scattering the phase difference between the excitations of
different nuclei is  exactly canceled by the phase produced by the path
difference in forward  propagation. This is independent of the structure and
the order of the  sample.

\item In an ordered crystal lattice, the excitation phases can be exactly
compensated at  Bragg angles to give constructive interference and an enhanced
peak.  However, at other angles the scattering is still coherent and the
interference is destructive, giving a smaller peak than would be obtained  if
the scattering were incoherent.

\item In a completely disordered crystal, the coherence of the incident
X-ray excitation is completely lost, since the outgoing phases are completely
random. Therefore the $90^o$ elastic peak will be stronger in a disordered
sample that in an ordered sample where there is destructive interference.

\item In samples where the distance between nearest neighbors is much smaller
than the photon wave length a coherent isotropic emission can occur. How much
of this emission is isotropic and how much is forward depends on the conditions
and parameters of the experiment. This includes the region
explored\cite{Scully} where $\vec k_0 \cdot \vec D$ is small but finite and
$ n\vec k_0 \cdot \vec D$ reaches unity for a finite value of $n \ll N$. This can
occur in quantum optics experiments but not in nuclear X-rays.

\end{enumerate}

There seems to be a crucial qualitative difference between the directed
spontaneous emission of optical and  X-ray photons from an extended  ensemble
of $N$ atoms. This has been shown in the discussion following eq.(\ref{geomet})
The first  paragraph of ref.(\cite{Scully}) raises two questions whose answers
may be very different in different areas.

\begin{enumerate}

\item Why don't we know which atom is excited? Simple ignorance will not give
coherence. There must be a quantum mechanical reason which prevents the
production mechanism and the emission of the emitted photon from leaving a
trail identifying the active atom\cite{nunowx}. This reason can be very
different in different areas of physics. A photon carries momentum and  the
momentum transfer identifies the active atom. The momentum transfer when a
free nucleus emits or absorbs a nuclear resonance photon is enormous; the
recoil energy transfer is many orders of magnitude larger than the natural line
width. The momentum transfer can be absorbed without a trace by a nucleus only
if it is localized to an extent where the Heisenberg momentum uncertainty is as
great as the momentum transfer. The physics of atomic optics is clearly very
different.

\item Why is directional correlation counterintuitive? Atomic physicists may
know  that a Dicke superradiant state will always emit a photon in 4$\pi$ sr.
if the atoms are all in a spatial region which is very much smaller than the
photon wave length. But X-ray crystallographers know that the photons in
coherent X-ray scattering scattering are highly correlated and that there is
coherent and incoherent scattering.

\end{enumerate}

\section{Previous history for X-ray scattering}

This issue has been already considered in detail on the pages of Physical
Review Letters\cite{HanTram,HJLIP,Ruffer,vanBurck} for the case of resonant
nuclear scattering. The correct answer  is certainly not counterintuitive.

The basic physics  needed to understand directed spontaneous emission from an
extended  ensemble of $N$ atoms goes back much further even than Dicke's
seminal paper on superradiance\cite{Super} or Ott's 1935 X-ray paper
\cite{OTT}, Lamb's 1949 paper\cite{LAMB} and Van Hove's 1954 paper\cite{Leon}.
It is a transition from an initial state of a system with many atoms in which
one atom is excited and we don't know which one and which has no photon to a
final  state in which all the atoms are in their ground state with no
excitation and there is one emitted photon wi<th wave vector $\vec k_0$.

This transition is described to lowest order in a perturbation series in the
fine structure constant  $\alpha$ by the Fermi ``golden rule" as proportional
to the square of the transition matrix element. There is no point to going back
to the elementary quantum mechanics needed to derive the Fermi golden rule. The
only problem is to define properly the initial and final states and express the
experimental observation in terms of sums of squares of the relevant matrix
elements.

The relation between Lamb's treatment\cite{LAMB} of neutron capture in crystals
and Ott's X-ray treatment\cite{OTT} was first pointed out by
Kaufman\cite{Bruria,HJLKauf} and reported in detail in a history of these
developments\cite{PHANNA}. A general formulation including these and other
processes of momentum transfer to bound systems is given in a quantum mechanics
book \cite{LipQM} which shows the relation of the dual wave-particle
descriptions of similar phenomena.

That scattered photons go in the direction of the absorbed photon is classical
wave optics and Huygens' principle, which states that radiation scattered from
an assembly of scatterers or emitted coherently from an assembly of sources are
described by combining the waves of all the scatterers or sources with the
proper phases. That the transmission of light through a medium is directionally
correlated with the direction of the incident beam is not counterintuitive at
all.

The N-atom state in a crystal prepared by the absorption of a single photon has
been well known in nuclear resonance physics and called a ``nuclear exciton" by
Hannon and Trammell\cite{HTREVIEW}. However its application is always in cases
where the distance between nearest neighbors is always at least of the order of
the photon wave length. A crystal never has a large number of nuclei confined to
a region much smaller than the wave length.

\section{Some further details}

\centerline {\bf {Angular Distributions - Basic physics often missed}}

\begin{enumerate}

\item  M\"ossbauer
 scattering from single nucleus  essentially isotropic.  Variations
 involve only lowest spherical harmonics.

\item Angular distribution of coherent forward
 scattering  sharply forward peaked.

\item  Forward peaking from  constructive interference
 between  amplitudes from different nuclei.

\item  Combination of constructive and destructive interference
at other angles

\begin{itemize} \item  Crucial to  understanding of all
experiments and  not generally understood.

\item Interference at angles far from forward; e.g. $90^o$
generally destructive.

\item Exact forward amplitude has zero solid angle
Experiments measure integral of intensity
over finite width of  forward peak.

\item Forward peak intensity and width both vary with thickness
of  sample, and both must be considered.

\end{itemize}
\end{enumerate}

\section*{Acknowledgments}

Some of these issues were clarified in discussions at the April, 2000 workshop
at HASYLAB\cite{HASYLAB}. It is a pleasure to thank E. E. Alp, A.Q.R.Baron,
U.van Buerck, R. Coussement H.Franz, E. Gerdau, W.Potzel, R.Roehlsberger,
W.Sturhahn and T.Toellner for helpful discussions and comments.

This work was supported
in part by the U.S. Department
of Energy, Basic Energy Sciences, Office of Science, under Contract
No.W-31-109-Eng-38.

%
\catcode`\@=11 
\def\references{
\ifpreprintsty \vskip 10ex
%
\hbox to\hsize{\hss \large \refname \hss }\else
\vskip 24pt \hrule width\hsize \relax \vskip 1.6cm \fi \list
{\@biblabel {\arabic {enumiv}}}
{\labelwidth \WidestRefLabelThusFar \labelsep 4pt \leftmargin \labelwidth
\advance \leftmargin \labelsep \ifdim \baselinestretch pt>1 pt
\parsep 4pt\relax \else \parsep 0pt\relax \fi \itemsep \parsep \usecounter
{enumiv}\let \p@enumiv \@empty \def \theenumiv {\arabic {enumiv}}}
\let \newblock \relax \sloppy
  \clubpenalty 4000\widowpenalty 4000 \sfcode `\.=1000\relax \ifpreprintsty
\else \small \fi}
\catcode`\@=12 

\end{document}